\documentclass[aps,pra,twocolumn,superscriptaddress]{revtex4-1}
\usepackage{graphicx}
\usepackage{amssymb}
\usepackage{amsmath}
\usepackage{bm}
\usepackage{xcolor}
\usepackage{verbatim}

\usepackage{hyperref}
\hypersetup{
     colorlinks=true,
     linkcolor=blue,
     filecolor=blue,
     citecolor=black,
     urlcolor=black,
     }
\usepackage[all]{hypcap}

\begin{document}

\title{Minimum critical velocity of a Gaussian obstacle in a Bose-Einstein condensate}

\author{Haneul Kwak}
\affiliation{Department of Physics and Astronomy, Seoul National University, Seoul 08826, Korea}

\author{Jong Heum Jung}
\affiliation{Department of Physics and Astronomy, Seoul National University, Seoul 08826, Korea}

\author{Y. Shin}
\email{yishin@snu.ac.kr}

\affiliation{Department of Physics and Astronomy, Seoul National University, Seoul 08826, Korea}

\affiliation{Center for Correlated Electron Systems, Institute for Basic Science, Seoul 08826, Korea}

\affiliation{Institute of Applied Physics, Seoul National University, Seoul 08826, Korea}

\begin{abstract}
When a superfluid flows past an obstacle, quantized vortices can be created in the wake above a certain critical velocity. In the experiment by Kwon {\it et al.} [Phys.~Rev.~A {\bf 91} 053615 (2015)], the critical velocity $v_c$ was measured for atomic Bose-Einstein condensates (BECs) using a moving repulsive Gaussian potential and $v_c$ was minimized when the potential height $V_0$ of the obstacle was close to the condensate chemical potential $\mu$. Here we numerically investigate the evolution of the critical vortex shedding in a two-dimensional BEC with increasing $V_0$ and show that the minimum $v_c$ at the critical strength $V_{0c}\approx \mu$ results from the local density reduction and vortex pinning effect of the repulsive obstacle. The spatial distribution of the superflow around the moving obstacle just below $v_c$ is examined. The particle density at the tip of the obstacle decreases as $V_0$ increases to $V_{c0}$ and at the critical strength, a vortex dipole is suddenly formed and dragged by the moving obstacle, indicating the onset of vortex pinning. The minimum $v_c$ exhibits power-law scaling with the obstacle size $\sigma$ as $v_c\sim \sigma^{-\gamma}$ with $\gamma\approx 1/2$.
\end{abstract}

\maketitle

\section{Introduction}
\label{sec:1}

A superfluid can flow without friction but only below a certain critical velocity. Above the critical velocity, the superfluid becomes dynamically unstable, generating excitations such as phonons and quantized vortices~\cite{Landau1941}. Understanding the critical dynamics and critical velocity of a superfluid is of fundamental and practical importance for the study of the transport properties of a superfluid system~\cite{Anderson1966,Blatter1994,Varoquaux2015}. The key questions are what induces the instability of the superfluid flow and how the energy dissipation evolves with the increasing flow velocity. At substantially high velocities, turbulent states would be developed in the superfluid system with a complex tangle of vortex lines, namely, quantum turbulence~\cite{Vinen2002,White2014}.

In recent experiments with atomic Bose-Einstein condensates (BECs), a localized optical potential formed by focusing a laser beam was adopted as a movable obstacle~\cite{Raman1999}. Various superfluid dynamics were investigated by controlling the movement of the obstacle in a sample. From the onset of energy dissipation with increasing obstacle speed, critical velocities of various atomic superfluid gases were demonstrated \cite{Raman1999,Dalibard12,2015_1,Moritz15,Park,Kim21}, where the measurement results tested theoretical predictions~\cite{Frisch1992,Jackson,Huepe,Rica,Crescimanno00,Zwerger00} and revealed the details of the dissipation mechanisms~\cite{2015_1,Moritz15,Park,Kim21,Singh2016,Singh2017}. For a fast obstacle above the critical velocity, the vortex shedding in the wake of the moving obstacle was investigated~\cite{2015_2, 2016, Lim, Von_Karman, Reeves}. A remarkable observation was that vortex clusters consisting of like-sign vortices are regularly shed from a uniformly moving obstacle in atomic BECs~\cite{2016}. This is analogous to the von Kármán vortex street in the classical viscous fluids in the transition to turbulence~\cite{Von_Karman,Reeves}.

For the optical obstacle, there are two regimes with respect to the relative magnitude of the obstacle's peak potential $V_0$ to the chemical potential $\mu$ of the BEC. The particle density at the obstacle position is suppressed because of the repulsion of the obstacle. However, when $V_0<\mu$, the condensate can penetrate the obstacle and a zero-density region is not induced in the condensate. In this penetrable case, vortices can be created only in the form of a dipole consisting of two vortices of opposite circulations. When $V_0>\mu$, which is referred to as impenetrable, a density-depleted hole is formed in the system, and it would significantly alter the characteristics of the vortex shedding dynamics by allowing the generation of vortex clusters~\cite{2016}. In the experiment by Kwon~{\it et al.}~\cite{2015_1}, the critical velocity $v_c$ for vortex shedding was measured as a function of $V_0$, and $v_c$ was minimized sharply at a certain critical strength $V_{0c}$ that was close to $\mu$. This implies that the onset behavior of the vortex shedding, which we refer to as critical vortex shedding, undergoes a certain transition as the obstacle strength changes from penetrable to impenetrable.

In this paper, we numerically study the critical vortex shedding of a Gaussian obstacle in a two-dimensional (2D) BEC and investigate its evolution with increasing obstacle strength. We verify that the critical velocity is minimized at a critical obstacle strength $V_{0c}$ close to $\mu$ and show that it arises from the start of vortex pinning as $V_0$ increases above $V_{0c}$. We examine the spatial distribution of the superflow around the moving obstacle just below $v_c$. At the critical strength, the superflow distribution suddenly changes to form a vortex dipole that is pinned at the tip of the obstacle. When $V_0$ is further increased, a density-depleted region develops and the co-moving, pinned vortex dipole becomes virtual and is absorbed in the region. The minimum $v_c$ at $V_0=V_{0c}$ decreases with increasing the obstacle size $\sigma$. We find that it exhibits a power-law scaling of $v_c\sim \sigma^{-\gamma}$, with $\gamma\approx 1/2$, which is in reasonable agreement with the experimental results of Ref.~\cite{2015_1}. Our results demonstrate the existence of the minimum critical velocity for a Gaussian obstacle and elucidate the transition of the critical vortex shedding from the penetrable to impenetrable regime.

The remainder of this paper is organized as follows. In Sec.~\ref{sec:2}, we describe a theoretical model to study the vortex shedding dynamics in a BEC based on the 2D Gross-Pitaevskii equation. In Sec.~\ref{sec:3}, we present numerical results, including a comparison of the shedding dynamics for penetrable and impenetrable obstacles, and the characterization of the critical vortex dipole state generated by the moving obstacle at the critical strength. Finally, in Sec.~\ref{sec:4}, a summary of this work and the outlooks for future studies are provided.

\section{Theoretical model}
\label{sec:2}

We consider a situation where an obstacle moves in a homogeneous BEC with a constant velocity $\textbf{v}$. In the mean-field theory, the BEC dynamics is described by the Gross-Pitaevskii equation (GPE),
\begin{equation}
    i \hbar\dfrac{\partial \Psi}{\partial t } =  \left(-\dfrac{\hbar^2}{2m} \nabla^2 + V(\textbf{r}-\textbf{v}t) + g|\Psi|^2 - \mu \right) \Psi, 
    \label{GPE}
\end{equation}
where $\Psi(\textbf{r},t)$ is the macroscopic wave function of the BEC, $\hbar$ is Planck's constant divided by $2\pi$, $m$ is the atom mass, $V(\textbf{r})$ is the obstacle potential, and $g$ is the nonlinear coupling coefficient. Taking the unitary transformation $\Psi(\textbf{r},t)=\exp [-\textbf{v}t\cdot \nabla ] \psi(\textbf{r},t)$, Eq.~(\ref{GPE}) is transformed into the reference frame moving with the obstacle as
\begin{equation}
    i \hbar\dfrac{\partial \psi}{\partial t } = \left(-\dfrac{\hbar^2}{2m} \nabla^2 + i\hbar v \dfrac{\partial}{\partial x}+ V(\textbf{r}) + g|\psi|^2 - \mu   \right) \psi
    \label{dGPE}
\end{equation}
with $\textbf{v}=v\hat{\textbf{x}}$. The characteristic length and time scales of the system are given by the healing length $\xi=\hbar/\sqrt{2m\mu}$ and $t_\mu=\hbar/\mu$, respectively. Using the change in variables, $\tilde{\textbf{r}}=\textbf{r}/\xi$ and $\tilde{t} =t/ t_\mu$, the equation can be expressed in a dimensionless form as
\begin{equation}
    i \partial_{\tilde{t}} \tilde{\psi} =   \left( - \tilde{\nabla}^2 + i \sqrt{2}\tilde{v} \partial_{\tilde{x}} + \tilde{V} (\tilde{\textbf{r}}) + |\tilde{\psi}|^2 - 1  \right) \tilde{\psi}
    \label{GPE_ND}
\end{equation}
with $\tilde{\psi}=n_0^{-1/2}\psi$, $\tilde{v}=v/c_s$, $\tilde{\nabla}=\xi \nabla$, and $\tilde{V}=V/\mu$. Here $n_0=\mu/g$ is the particle density of the BEC without the obstacle and $c_s=\sqrt{\mu/m}$ is the speed of sound. 

In this work, we study the BEC dynamics for a Gaussian obstacle in two dimensions. This is motivated by the recent experiments using highly oblate atomic samples~\cite{2015_1,Moritz15,Park,Kim21,Lim}, where the vortex line dynamics along the tight confining direction is energetically irrelevant. Hence, the shedding dynamics can be well described in 2D. In a hydrodynamic approximation, the dimensional reduction is carried out by integrating the wave function component along the short axis. It effectively modifies the speed of sound in Eq.~(\ref{GPE_ND})~\cite{Stringari,Kim20}. The potential of the Gaussian obstacle is given by $V(r)=V_0 \exp[-2(r^2/\sigma^2)]$, where $r=\sqrt{x^2+y^2}$ and $\sigma$ is the $1/e^2$ radius of the obstacle. The obstacle is located at the origin of the reference frame.

We numerically solve Eq.~(\ref{GPE_ND}) in the $xy$ plane with periodic boundary conditions, using the pseudo spectral method~\cite{GPELab_2}. In the simulation of vortex shedding for $v>v_c$, we set the initial state to be a stationary solution for a velocity $v_i$ slightly below $v_c$. Next, we increase the obstacle speed up to the target velocity $v$ for an acceleration time $t_{a}=200t_{\mu}$~\cite{footnote1}. The initial stationary solution is obtained using the imaginary-time method where $t$ is replaced by $-i\tau$~\cite{Dalfovo96,imaginary_time}. To realize a constant stream at the front boundary of the obstacle, we adopt the numerical method described in Ref.~\cite{Reeves}, where damping zones with a thickness of $20\xi$ are set at the boundary to attenuate the wake of the BEC and recover the constant uniform flow at the front boundary. In the calculation of a stationary solution using the imaginary-time propagation method, the damping zone is inactivated.


\section{Results and discussion}
\label{sec:3}

\subsection{Determination of critical velocity}

\begin{figure}[!t]
    \centering
    \includegraphics[width=\linewidth]{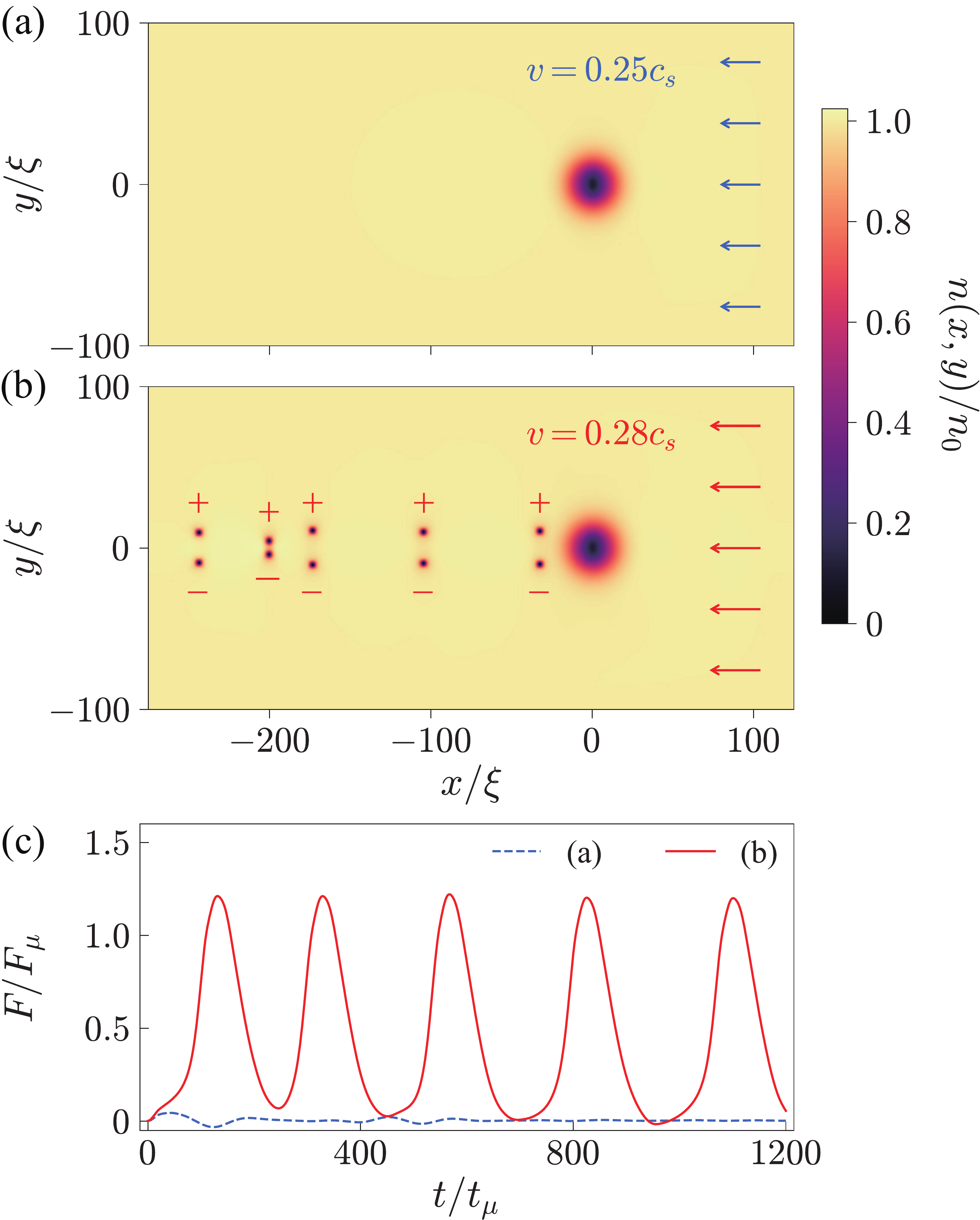}
    \caption{Vortex shedding from a Gaussian obstacle in a Bose-Einstein condensate. Particle density distribution $n(x,y)$ of a BEC flowing past an obstacle with size $\sigma/\xi=20$ and strength $V_0/\mu=0.8$ for flow speed (a) $v/c_s=0.25$ and (b) $0.28$, at time $t/t_{\mu}=1200$. $n_0$ denotes the particle density of the BEC without the obstacle and the flow direction is indicated by the arrows. In (b), the flow speed is faster than the critical velocity of $v_c\approx 0.26 c_s$, and vortices are generated behind the obstacle. The circulation directions of the vortices are indicated by $+$ (counterclockwise) and $-$ (clockwise). (c) Temporal evolution of the drag force $F$ experienced by the BEC in (a) and (b). $F_{\mu}=\mu/\xi$.}
    \label{fig:FIG1}
\end{figure}


Figures~\hyperref[fig:FIG1]{\ref{fig:FIG1}(a)} and \hyperref[fig:FIG1]{\ref{fig:FIG1}(b)} display the density distributions of the BEC, $n(x,y)=|\psi|^2$, at $t/t_{\mu}=1200$ for two different velocities, $v/c_s=0.25$ and $0.28$, respectively~\cite{footnote1}. The obstacle size and strength are $\sigma/\xi=20$ and $V_0/\mu=0.8$. When the speed is lower than the threshold value of $v_c\approx 0.26 c_s$, no vortices are generated. The BEC remains stationary [Fig.~\hyperref[fig:FIG1]{\ref{fig:FIG1}(a)}]. By contrast, when the obstacle velocity increases above the threshold velocity, vortices are emitted from the obstacle in a periodic manner~\cite{2015_2}. The periodic vortex shedding is also examined by inspecting the drag force exerted by the obstacle $F_x=-\int\tilde{\psi}^*(\partial_{\tilde{x}} \tilde{V})\tilde{\psi} d^2\tilde{\textbf{r}}$ using the Ehrenfest relation~\cite{Reeves}. We verify that for $v>v_c$ the force oscillates in time, corresponding to the periodic vortex emission. For $v<v_c$ it is stationary and remains approximately zero [Fig. \hyperref[fig:FIG1]{\ref{fig:FIG1}(c)}].

We determine the critical velocity $v_c$ from the existence of a stationary ground state solution via the imaginary time propagation method~\cite{Huepe}. The imaginary time method gives a converging stationary solution for $v<v_c$ or an oscillating solution otherwise. In the oscillating solution, a pair of vortices is created by the obstacle. They move away from each other along the $y$-direction and are annihilated at the system's boundary due to the periodic boundary conditions. This process is repeated over an imaginary time. In the calculation of stationary solutions, we employed a spatial domain of $(L_x,L_y)=(400,400)\xi$ with $(N_x,N_y)=(600,600)$ grids and took a time step of $\Delta \tau/t_{\mu}=0.04$. We decided the convergence of a solution through its behavior up to the imaginary time $\tau/t_{\mu}=4000$. The critical velocities determined from our imaginary time method are identical to the threshold values from the simulation of the real-time evolution within an error of $0.02c_s$.

\begin{figure}[!t]
    \centering
    \includegraphics[width=0.95\linewidth]{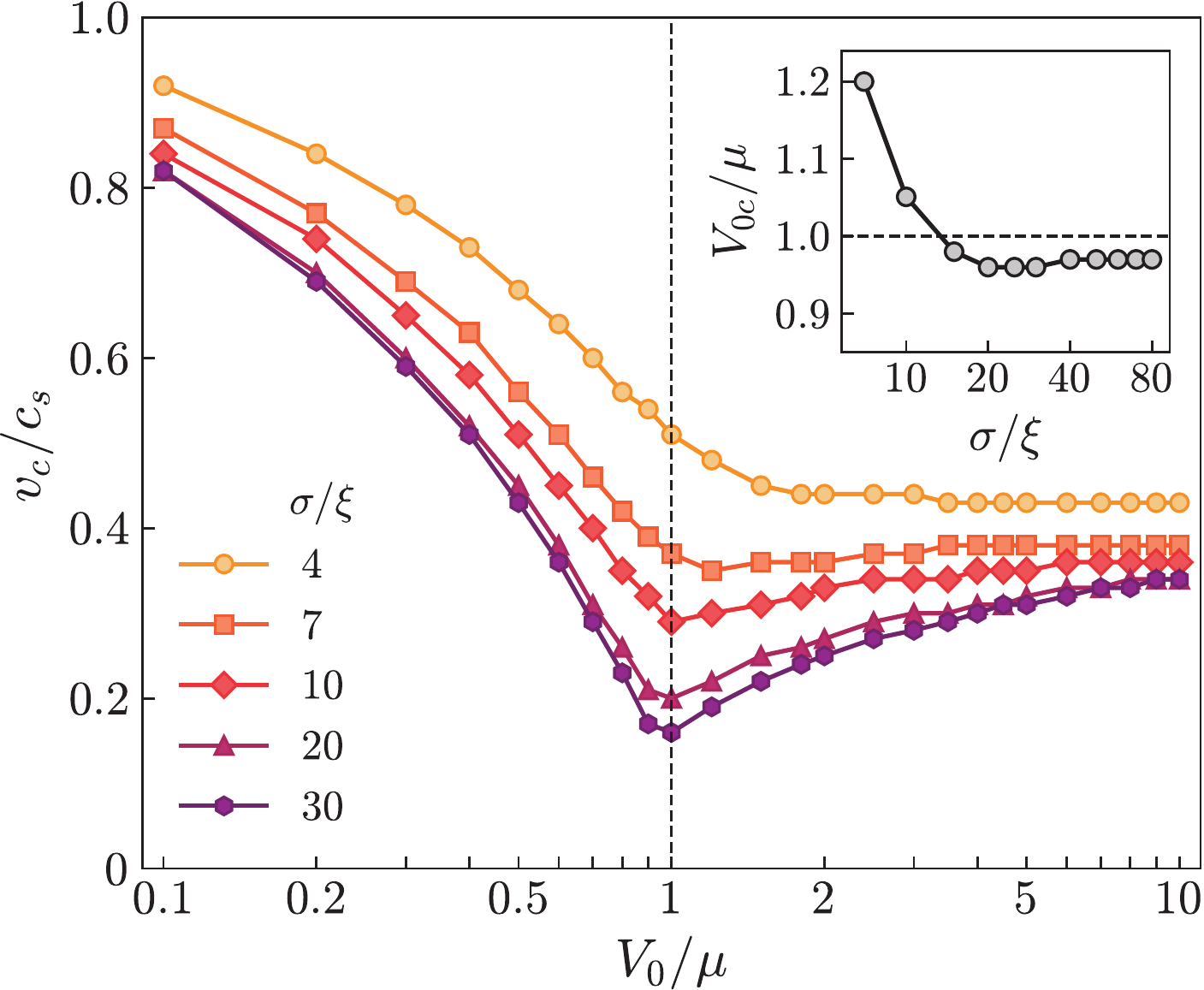}
    \caption{Critical velocity $v_c$ as a function of the obstacle strength $V_0$ for various obstacle sizes $\sigma$. The $x$-axis has a logarithmic scale. For $\sigma/\xi>4$, $v_c$ is minimized at a critical strength $V_{0c}$, close to the chemical potential $\mu$ of the BEC. The dashed line denotes $V_0 = \mu$ and represents the boundary between the penetrable regime and the impenetrable regime. The inset shows the critical obstacle strength $V_{0c}$ as a function of $\sigma$.}
    \label{fig:FIG2}
\end{figure}

\begin{figure*}[!t]
    \centering
    \includegraphics[width=\linewidth]{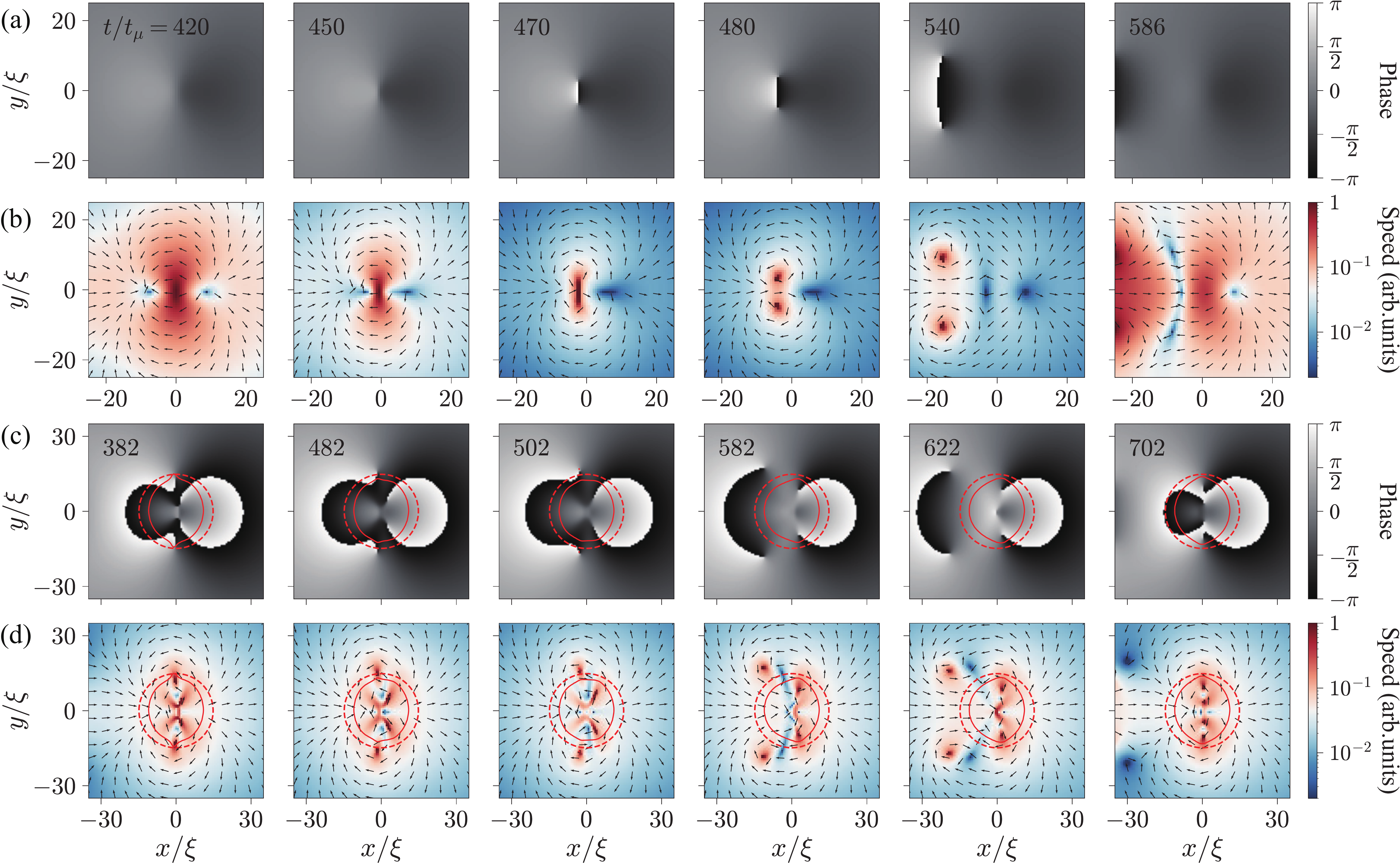}
    \caption{Vortex nucleation process. Time evolution of the phase (a,c) and velocity field (b,d) of a BEC for a penetrable obstacle ($\sigma/\xi=20,\,V_0 /\mu=0.8,\,v/c_s=0.27$) in (a) and (b), and for an impenetrable obstacle ($\sigma/\xi=20,\,V_0 /\mu=3.0,\,v/c_s=0.30$) in (c) and (d). In (b) and (d), the flow speed is normalized with the maximum speed in each panel. In (c) and (d), the red dashed circles delineate the boundary of the region where the obstacle's repulsive potential is larger than the chemical potential $\mu$ of the BEC. The red solid lines indicate the particle-density contour line at $0.1\%$ of $n_0$. The velocity field in the zero-density region is virtual.}
    \label{fig:FIG3}
\end{figure*}

Figure~\ref{fig:FIG2} displays the numerical results of the critical velocities over a range of obstacle strength $0.1\le V_0/\mu\le 10$ for various obstacle radii $4\le\sigma/\xi\le30$. We observe a marked dip of $v_c$ with a minimum around $V_0=\mu$, which agrees well with the previous experimental observation~\cite{2015_1}. In the limit of $V_0\rightarrow0$, $v_c$ approaches to the speed of sound $c_s$. It is compatible with the fact that the critical velocity of a microscopically small impurity is given by the speed of sound according to the Landau criterion, although such a small obstacle would generate phonons or a localized rarefaction pulse called the Jones-Roberts soliton~\cite{soliton,Meyer2017}, rather than vortices. In the limit of $V_0\rightarrow\infty$, the Gaussian obstacle is well represented as a hard cylinder. $v_c$ is expected to converge to a constant value of approximately $0.4 c_s$. It was verified numerically~\cite{Huepe} and analytically~\cite{Rica} that the critical velocity of a large hard cylinder is given by $\approx 0.37 c_s$ regardless of the diameter of the cylinder. 

The local Landau criterion provides a qualitative interpretation of the observed $V_0$ dependence of $v_c$. Namely, when $V_0<\mu$, the particle density in the obstacle region decreases with an increasing $V_0$ as $n\approx (\mu-V_0)/g$. This lowers the local speed of sound at the tip of the obstacle and consequently, decreases the critical velocity. However, when $V_0>\mu$, a density-depleted region is formed by the obstacle. Vortices would be generated at the flanks of the obstacle, restoring $v_c$ to that in the hard cylinder case. However, it is not clear why the critical velocity sharply changes its behavior at the critical obstacle strength $V_{0c}\approx \mu$.
In addition, when the obstacle size is reduced below $10\xi$, the critical obstacle strength of the minimum $v_c$ is slightly shifted to a higher $V_0$. Eventually, for $\sigma/\xi<7$ the local minimum of $v_c$ does not occur in our investigation range of $V_0$. The inset in Fig.~\ref{fig:FIG2} shows $V_{c0}$ as a function of $\sigma$. The main purpose of this work is to probe the underlying mechanism of the minimum $v_c$ at the critical obstacle strength.

\subsection{Penetrable-to-impenetrable transition}

We first compare the characteristics of the vortex generation dynamics for penetrable and impenetrable obstacles. In Fig.~\ref{fig:FIG3}, we display a time sequence of the phase and velocity field distributions around the obstacle as vortices are generated for the two cases with $V_0/\mu=0.8$ and $3.0$, respectively. The velocity field of a superfluid is determined from the probability current $\textbf{j}=-\frac{i\hbar}{2m}(\psi^*\nabla\psi-\psi\nabla\psi^*)=\frac{\hbar}{m}|\psi|^2\nabla\Phi\equiv n \textbf{v}_s$ with $\Phi(\textbf{r})$ being the phase of the macroscopic wave function $\psi(\textbf{r})$.

\begin{figure*}[!t]
    \centering
    \includegraphics[width=\textwidth]{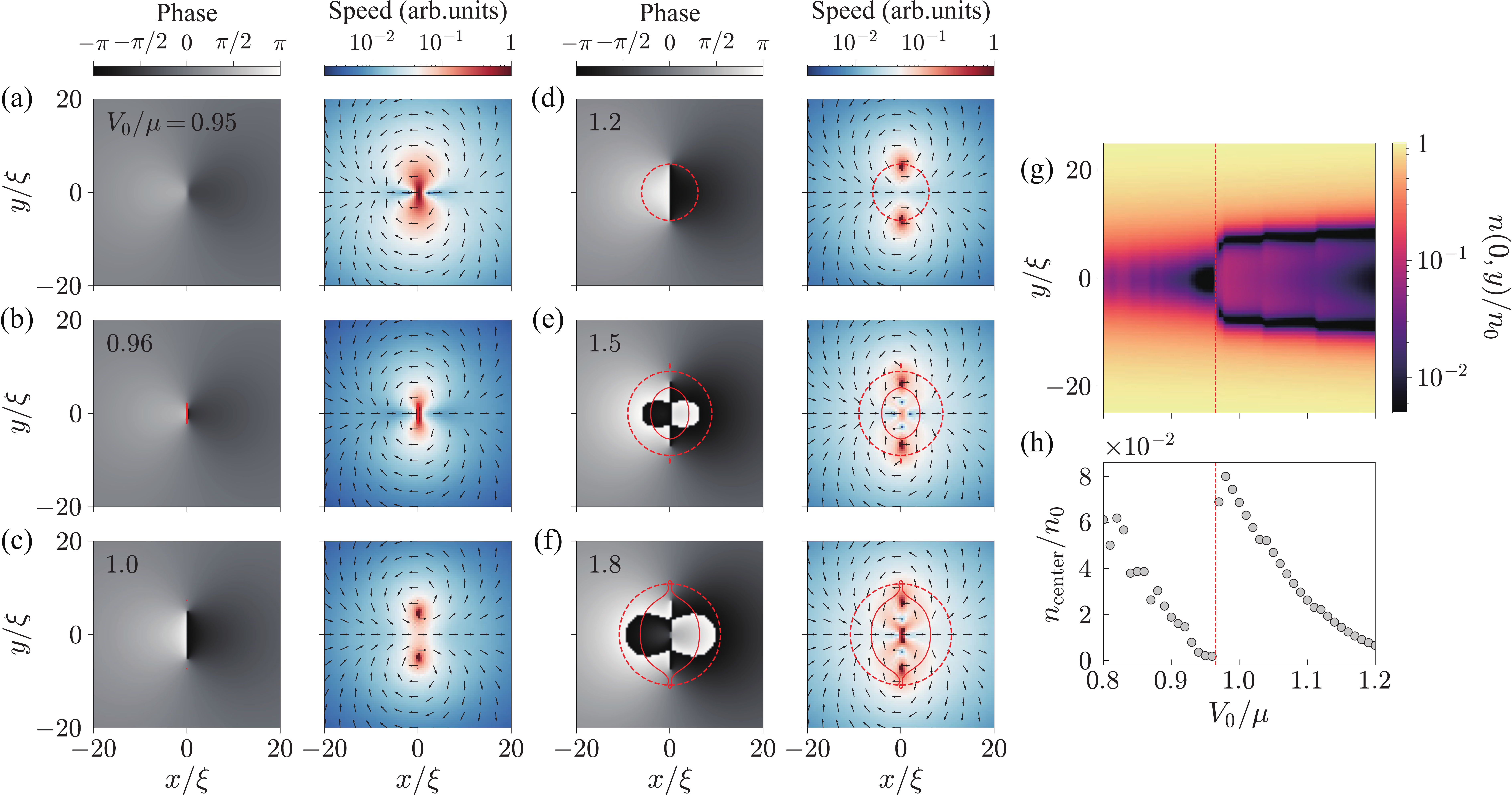}
    \caption{Penetrable-to-impenetrable transition of the critical superflow around the moving Gaussian obstacle. (a--f) Spatial distributions of the phase (left) and velocity field (right) of a BEC for various obstacle strengths (a) $V_0/\mu=0.95$, (b) 0.96, (c) 1.0, (d) 1.2, (e) 1.5, and (f) 1.8, where $\sigma/\xi=20$. The obstacle velocity is $v\rightarrow v_c^-$, i.e., just below $v_c$ for each obstacle strength. The colormap and red dashed and solid lines have the same meaning as in Fig.~\ref{fig:FIG3}. (g) Particle density profile $n(0,y)$ of the BEC at the critical velocity as a function of $V_0$. The colorbar has a logarithmic scale. Just above the critical strength $V_{0c}=0.96 \mu$, the density minimum is bifurcated, corresponding to the vortex dipole formation. (h) The density $n_\textrm{center}$ at the center of the obstacle is shown as a function of $V_0$, which corresponds to the horizontal line of $y=0$ in (g). The red dashed lines in (g,h) mark the critical strength $V_{0c}$.}
    \label{fig:FIG4}
\end{figure*}

For the penetrable obstacle, two zero-velocity regions, indicated by the dark blue area in Fig. \ref{fig:FIG3}(b), are formed at the front and rear of the obstacle, respectively, and they have a high velocity region between them. The mass flow diverges out in the front region and converges behind the obstacle. This is a consequence of the pressure increase at the front and the subsequent pressure drop behind the moving obstacle. As time passes, the rear zero-velocity region gradually disappears and the high-velocity area evolves into a vortex pair separated along the $y$ direction. At the vortex nucleation moment, the flow direction in the high-velocity region is rapidly flipped due to the phase accumulation and slippage~\cite{Jackson,footnote2}. As the vortex dipole is shed, the front zero-velocity region is separated into two parts along the obstacle's moving direction. The same vortex dipole generation process is repeated.

For the impenetrable obstacle case of $V_0/\mu=3$, two pairs of vortices exist inside a density-depleted region [Figs. \hyperref[fig:FIG3]{\ref{fig:FIG3}(c)} and \hyperref[fig:FIG3]{\ref{fig:FIG3}(d)}]. These virtual vortices correspond to the superflow pattern around the obstacle. Vortex emission occurs by peeling off the two outer vortices into the condensate. The remaining two vortices inside the density-depleted zone come out to the obstacle's boundary and simultaneously, a new vortex pair is produced at the center of the obstacle. It forms a configuration identical to that before the vortex emission, thus preparing for the next emission. This vortex emission via unpinning from the obstacle's boundary layer is qualitatively different from that for the penetrable obstacle, where a vortex dipole is generated via phase slippage at the tip of the obstacle.

To investigate the evolution of the critical vortex shedding with an increasing $V_0$, in Fig.~\hyperref[fig:FIG4]{\ref{fig:FIG4}} we display the phase and velocity fields of the stationary solutions at $v\rightarrow v_c^-$, for $\sigma/\xi=20$ and various $V_0$. In the penetrable regime, as $V_0$ increases to $V_{0c}$, the two zero-velocity regions get closer and the in-between high-velocity region becomes more localized. At the critical potential strength $V_{0c}\approx\mu$, the obstacle creates phase discontinuity [Fig.~\hyperref[fig:FIG4]{\ref{fig:FIG4}(b)}]. Subsequently, a pair of vortices, which are pinned, i.e., stationary with the obstacle, appear [Fig.~\hyperref[fig:FIG4]{\ref{fig:FIG4}(c)}]. A density-depleted region is not formed yet at the center of the obstacle even for $V_0/\mu=1.2$ [Fig.~\hyperref[fig:FIG4]{\ref{fig:FIG4}(d)}]. As $V_0$ increases further, a zero-density region develops and  eventually, for $V_0/\mu>1.5$, the pinned vortex dipole is absorbed in the density-depleted region and becomes virtual [Fig.~\hyperref[fig:FIG4]{\ref{fig:FIG4}(e)}]. For even higher $V_0$, a new bound virtual vortex pair appears in the zero-density region [Fig.~\hyperref[fig:FIG4]{\ref{fig:FIG4}(f)}]. The evolution of the superflow pattern at the critical condition shows that the sudden formation of a vortex dipole pinned at the tip of the obstacle represents the transition of the critical vortex shedding from the penetrable to impenetrable regime.

The transition at the critical strength is also demonstrated in the evolution of the particle density profile along $x=0$ [Fig.~\hyperref[fig:FIG4]{\ref{fig:FIG4}(g)}]. As the obstacle strength increases in the penetrable regime, the density $n_{\textrm{center}}$ at the obstacle's tip decreases to zero as $V_0$ approaches $V_{0c}$. When $V_0$ exceeds $V_{0c}$, the BEC does not develop a density-depleted region at the tip of the obstacle. Instead, it creates a vortex pair, which is indicated by a bifurcation of the zero-density point in the density profile. Correspondingly, at the tip of the obstacle, $n_{\textrm{center}}$ shows a sudden jump [Fig.~\hyperref[fig:FIG4]{\ref{fig:FIG4}(h)}]. As a result, even if the obstacle height is higher than the chemical potential, $n_{\textrm{center}}$ remains finite.

The formation of a vortex dipole at $V_0>V_{0c}$ indicates the onset of the pinning effect from the repulsive obstacle. The pinning effect is checked by the fact that the bare linear velocity of the vortex pair, which is given by $v_d=\frac{\hbar}{m d}$ with separation $d$ \cite{Snell}, is smaller than the obstacle velocity. The Magnus force exerted on the vortices points outside of the obstacle~\cite{Groszek18}. Thus, the vortex dipole is dragged by the moving obstacle under the pinning. The vortex pinning effect makes vortex shedding difficult. Therefore, we attribute the sudden increase of $v_c$ when $V_0$ increases above $V_{0c}$ to the activation of the vortex pinning effect. Recently, in Ref. \cite{Reeves2}, the vortex pinning mechanisms were numerically investigated for a circular, uniform potential, and a similar stationary solution was reported.

The onset of the vortex pinning effect explains the observation in Fig. \ref{fig:FIG2} that for small obstacles, $V_{0c}$ is shifted towards a higher value and it disappears at $\sigma/\xi<7$. Due to the finite size of a vortex core, for sufficiently small obstacles, a vortex dipole with separate cores could not be stably produced in a pinned configuration and thus, $v_c$ monotonically decreases with increasing $V_0$. We label such a small obstacle as a \textit{quantum obstacle}~\cite{Reeves}. In general, strong impenetrable obstacles would generate vortex clusters, which consist of many same-sign vortices, for high velocity. However, lacking the pinning effect, a quantum obstacle would not lead to large vortex cluster shedding even for high $V_0$~\cite{Reeves}.  Some numerical results are presented in the Appendix.

\begin{figure}
    \centering
    \includegraphics[width=0.95\linewidth]{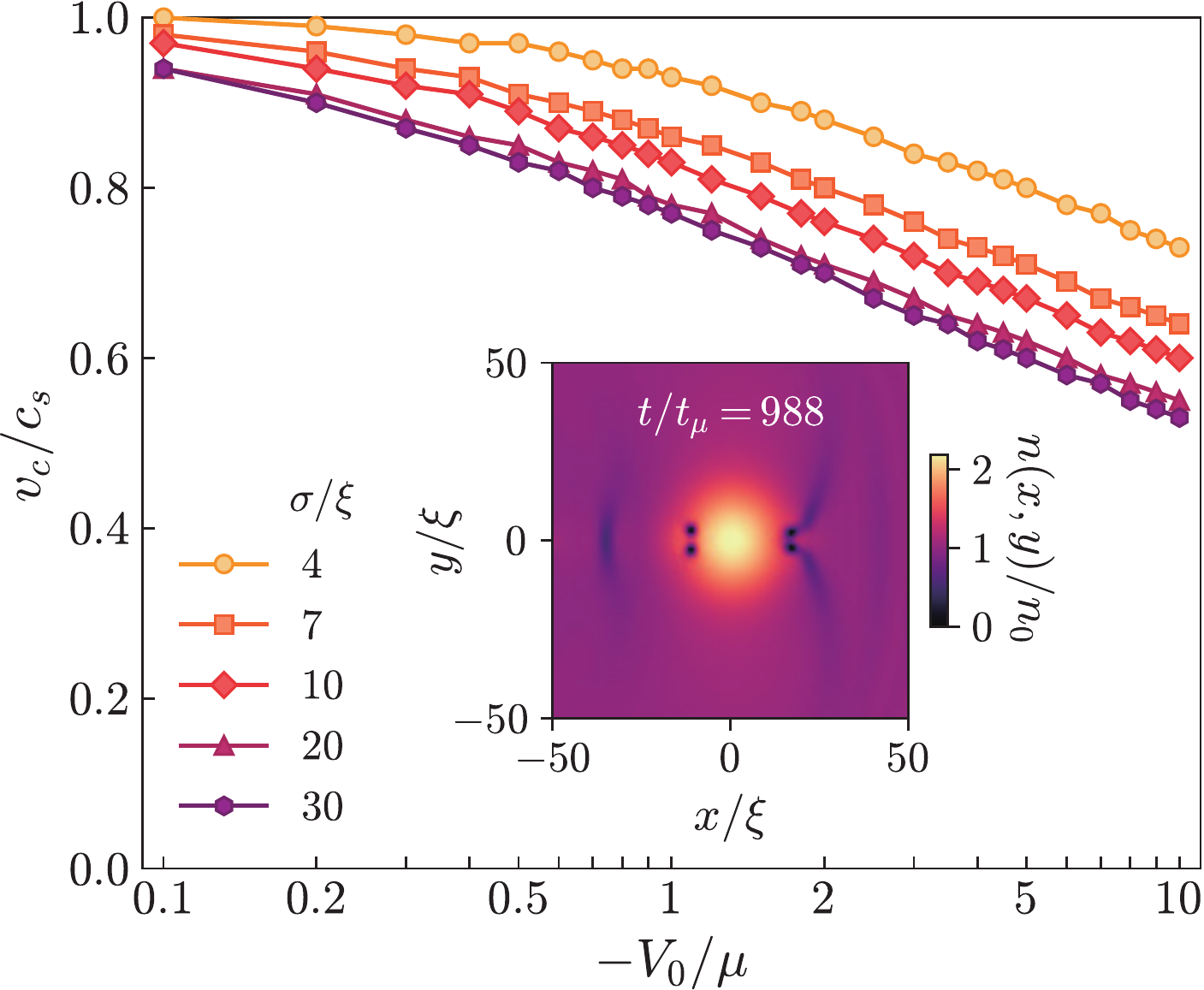}
    \caption{Critical velocities for various attractive obstacles. The $x$-axis has a logarithmic scale. The inset shows the particle density distribution $n(x,y)$ of a BEC flowing past an attractive obstacle with $\sigma/\xi=20$ and $V_0/\mu=-1$, where the obstacle is \textcolor{magenta}{is} accelerated from $0.7c_s$ to $0.85c_s$ for a time of $1000t_\mu$.}
    \label{fig:FIG5_attractive}
\end{figure}

Finally, for completeness, we calculate the critical velocities for attractive obstacles with negative $V_0$ and present the results in Fig.~\ref{fig:FIG5_attractive}. As $|V_0|$ increases, the critical velocity monotonically decreases. The inset in Fig.~\ref{fig:FIG5_attractive} shows the vortex creation process from an attractive obstacle, where a vortex dipole is generated from a rarefaction pulse being broken in front of the obstacle~\cite{Jackson2000, Aioi2011}.

\subsection{Obstacle size dependence}

In the study of the critical velocity of superflow past an obstacle, the dependence of $v_c$ on the obstacle size has attracted interest because it may reflect the dispersion property of the excitation mode involved in the critical energy dissipation~\cite{Feynman55,Zwerger00}. In our situation with a Gaussian obstacle, we observe that the variations of $v_c$ around $V_0=\mu$ becomes more pronounced for a larger $\sigma$ with lowering the minimum $v_c$ (Fig.~\hyperref[fig:FIG2]{\ref{fig:FIG2}}).

\begin{figure}[!t]
    \centering
    \includegraphics[width=\linewidth]{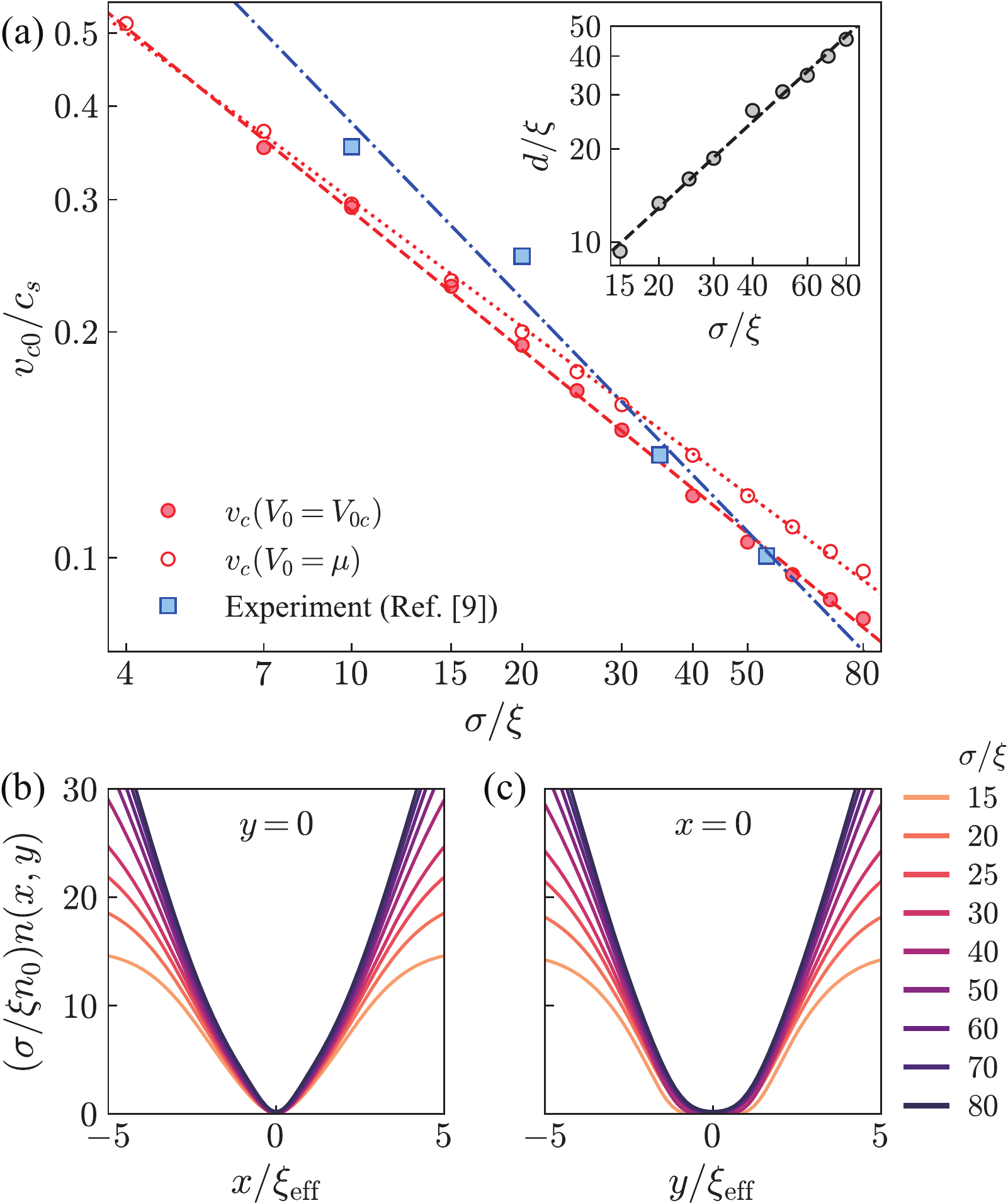}
    \caption{Obstacle size dependence of the critical velocity. (a) Minimum critical velocity $v_{c0}$ (red solid circles) and $v_c$ at $V_0=\mu$ (red open circles) as functions of $\sigma$. The blue squares are the experimental results from Ref.~\cite{2015_1}. All data are plotted in log-log scale. The lines are power-law functions of $v_c=v_0 {(\sigma/\xi)}^{-\gamma}$, fitted to the data sets with $\{v_0/c_s,\gamma\}=\{1.19,0.61\}$ for the minimum $v_c$, $\{1.09,0.56\}$ for $v_c$ at $V_0=\mu$, and $\{2.3,0.78\}$ for the experimental data. The inset displays the pair separation $d$ of the critical vortex dipole as a function of $\sigma$ at $V_0=\mu$. The black dashed line is a power-law function fit to the data, yielding $d=d_{0}(\sigma/\xi)^{\gamma_d}$ with $d_0=0.81\xi$ and $\gamma_d=0.92$. Rescaled particle density profiles along (b) $y=0$ and (c) $x=0$ at 
    $v\rightarrow v_c^{-}$ and $V_0\rightarrow V_{0c}^{-}$ for various obstacle sizes.}
\label{fig:FIG5}
\end{figure}

Figure~\hyperref[fig:FIG5]{\ref{fig:FIG5}(a)} displays the minimum critical velocity $v_{c0}$ at $V_0=V_{0c}$ as a function of the obstacle radius $\sigma$ in a log-log scale, together with the experimental measurement data from Ref.~\cite{2015_1}. First, our numerical results are in good quantitative agreement with the experimental results~\cite{footnote3}. Second, they suggest a power-law dependence of $v_{c0}$ on $\sigma$. From a power-law function of $v_{c0} = v_0 {(\sigma/\xi)}^{-\gamma}$ fitted to the data points, we obtain $\{v_0/c_s,\gamma\}=\{1.19(3), 0.61(1)\}$ for the numerical results and $\{2.3(7), 0.78(9)\}$ for the experimental data.

To understand the origin of the scaling behavior, we consider the GPE for $V_0=\mu$ in the large obstacle limit, $\sigma\gg\xi$, and in the obstacle center region, where the obstacle potential is approximated as ${V}({\textbf{r}}) \approx \mu[1-2(r/\sigma)]^2$. Expecting a scaling behavior of the critical superflow state with $\sigma$, we introduce new length and time scales as $\xi_{\textrm{eff}} = \xi s^{\alpha}$ and $t_{\mu,\textrm{eff}} = t_{\mu} s^{\beta}$ with $s=\sigma/\xi$, respectively. By changing the variables, $\bar{\textbf{r}} = {\textbf{r}}/\xi_{\textrm{eff}}$ and $\bar{t} = {t}/t_{\mu,\textrm{eff}}$, we obtain a dimensionless expression for the GPE, which is explicitly independent of $\sigma$. When $\alpha=\frac{1}{2}$ and $\beta=1$, Eq.~\eqref{dGPE} is recaptured in a $\sigma$-independent form as
\begin{equation}
    i \partial_{\bar{t}} \bar{\psi} =  \left( - \bar{\nabla}^2 + i\sqrt{2}\bar{v} \partial_{\bar{x}}-2\bar{r}^2 + |\bar{\psi}|^2 \right) \bar{\psi},
    \label{GPE_ND_NS_final}
\end{equation}
where $\bar{v}={s}^{1/2}(v/c_s)$, $\bar{\nabla} = \xi_{\textrm{eff}} \nabla$, and $\bar{\psi}=(n_0/s)^{-1/2}\psi$. This suggests that $v_{c}\propto 1/\sqrt{\sigma}$, which is close to the observed scaling behavior of the minimum critical velocity $v_{c0}$ with $\sigma$. Noting that $V_{0c}$ is not exactly equal to $\mu$, in Fig.~\hyperref[fig:FIG5]{\ref{fig:FIG5}(a)}, we also plot $v_c$ at $V_0=\mu$ (red open circles) as a function of $\sigma$. A power-law fitting gives $\{v_0/c_s,\gamma\}=\{1.09(2),0.56(1)\}$. It agrees better with the prediction of Eq.~\eqref{GPE_ND_NS_final}. The power-law relation estimates the critical velocity with considerable accuracy throughout the whole range of $\sigma$ including the quantum obstacle regime.

The dimensionless GPE of Eq.~\eqref{GPE_ND_NS_final} suggests the characteristic length scale of the system $\xi_{\textrm{eff}}=\sqrt{\xi \sigma}$. In Figs.~\hyperref[fig:FIG5]{\ref{fig:FIG5}(b)} and \hyperref[fig:FIG5]{\ref{fig:FIG5}(c)}, we plot the profiles of normalized particle density $\bar{n}=|\bar{\psi}|^2= \frac{1}{n_0}\frac{\xi_\textrm{eff}^2}{\xi^2}|\psi|^2$ along $y=0$ and $x=0$, respectively, for various $\sigma$, where the wave functions $\psi(\textbf{r})$ for the critical condition of $V_0 \rightarrow V_{0c}^{-}$ and $v \rightarrow v_{c0}^{-}$ are calculated from Eq.~(\ref{dGPE}). The normalized number densities collapse remarkably well in the center region of $r\ll\sigma$ as expected from Eq.~\eqref{GPE_ND_NS_final}. The length scale $\xi_{\textrm{eff}}$ may be regarded as an effective healing length for the average condensate density $\langle n\rangle$ in the obstacle center region, i.e., $\xi_{\textrm{eff}} = \hbar/\sqrt{2mg\langle n\rangle}$. Here,
$\langle n \rangle$ is estimated in a self-consistent manner as $\langle n \rangle = \frac{1}{\pi\xi_{\textrm{eff}}^2} \int_{r<\xi_{\textrm{eff}}} n(\textbf{r}) d^2 \textbf{r}$. It is the mean value over a disk area of radius $\xi_\textrm{eff}$. With $n(r)= n_0 \frac{2r^2}{\sigma^2}$ for $r\ll \sigma$, $\langle n\rangle= n_0 \frac{\xi_\textrm{eff}^2}{\sigma^2}$, yielding $\xi_{\textrm{eff}}=\sqrt{\xi\sigma}$.

Finally, we investigate the dependence of the pair separation $d$ of the critical vortex dipole state on the obstacle size. In the inset of Fig.~\hyperref[fig:FIG5]{\ref{fig:FIG5}(a)}, the pair separation $d$ at $V_0=\mu$ is plotted as a function of $\sigma$ in a log-log scale, and we observe that $d$ exhibits power-law scaling with $\sigma$ as $d\propto \sigma^{0.9}$. The scaling exponent is not accounted for by the length scale $\xi_\textrm{eff}$, which is understandable because the vortex separation is the order of the obstacle radius, invalidating the center-region approximation in Eq.~\eqref{GPE_ND_NS_final}.

According to Ref.~\cite{Reeves2,Groszek18}, the velocity of a vortex is given by
\begin{equation}
    \textbf{v}_v = \dfrac{\hbar}{m} \left( \nabla\Phi - \dfrac{1}{2}\hat{\bm{\kappa}} \times \nabla \ln n \right)\bigg|_{\textbf{r}_v},
    \label{Eq5}
\end{equation}
where $\textbf{r}_v$ denotes the vortex position and $\hat{\bm{\kappa}}$ is the circulation direction of the vortex. The first term describes a velocity from the ambient phase gradient without the vortex's singular contribution. The second term describes the one induced from the density gradient. In the critical vortex dipole state, assuming that the vortices generate a phase gradient equal to that in a homogeneous BEC, the first term arising from the counterpart vortex in the dipole is estimated as $\frac{\hbar}{m}\nabla \Phi \sim  \frac{\hbar}{md} \hat{\textbf{x}}$. Taking $n(r) = n_0 (1-e^{-2\frac{r^2}{\sigma^2}})$ in the Thomas-Fermi approximation, the second term gives $v_d= -\frac{\hbar}{m}\hat{\textbf{z}} \times \nabla \ln n = \frac{\hbar}{m} \frac{n_0-n(\frac{d}{2})}{n(\frac{d}{2})}  \frac{d}{\sigma^2}\hat{\textbf{x}}$. Then, from $\textbf{v}_v=v_c \hat{\textbf{x}}$, the observation of $d \sim \sigma$ suggests $v_c \sim 1/\sigma$, which is not compatible with the observed scaling behavior of $v_c$. This implies that the velocity field around the vortex dipole is significantly modified in the inhomogeneous density distribution due to the obstacle potential~\cite{inhomogeneous_vortex}. The structure and stability of the critical vortex dipole state are an interesting subject and warrants further investigation in future.

\section{Summary and outlook}
\label{sec:4}

We numerically investigated the critical velocity of a Gaussian obstacle in a uniform 2D BEC using the GPE. From the existence of a stationary solution, we determined the critical velocity as a function of the obstacle strength. It is minimized at the critical strength $V_{0c}\approx \mu$, which is consistent with the previous experimental results in Ref.~\cite{2015_1}. We examined the flow pattern of the condensate around the obstacle for the velocity just below $v_c$. A vortex dipole is abruptly formed at the tip of the obstacle as the obstacle strength exceeds $V_{0c}$. This sudden change in the critical flow pattern indicates the onset of the vortex pinning effect by the moving obstacle. It represents the penetrable-to-impenetrable transition of the vortex shedding dynamics. Further, the minimum critical velocity at the critical obstacle strength exhibits a power-law dependence on the obstacle size as $v_c\propto \sigma^{-\gamma}$ with $\gamma\approx 1/2$. Additionally, the measured exponent is explained by the scaling property of the GPE near the center of the obstacle with $V_0=\mu$. 

The superflow state where a vortex dipole is pinned and dragged by the moving obstacle presents an interesting situation for the study of the critical vortex shedding. As mentioned in the discussion of the $\sigma$ dependencies of $v_c$ and $d$, the structure of the critical vortex dipole state and its stability need to be further investigated. A force balance analysis including the Magnus force and vortex attraction in the background with inhomogeneous density might be fruitful~\cite{Groszek18,inhomogeneous_vortex,Thompson}. Near the critical shedding condition, small breathing motions of the vortex dipole were observed in our numerical simulation, where acoustic radiation from the vortex dipole-obstacle interaction is anticipated~\cite{Suthar14}. At the critical obstacle strength, the transition of the superflow distribution appears very rapid, similar to the first-order. Thus, it suggests that there might be some hysteresis effects in the vortex shedding when the obstacle changes its strength in time. In Ref.~\cite{hysteresis}, the bistability in the vortex shedding near the critical velocity was reported. Lastly, the recent experimental work on the vortex shedding frequency $f_v$ showed that the increasing rate of $f_v$ with the obstacle velocity is fastest at $V_0\approx \mu$~\cite{Lim}. This might have originated from the critical vortex dipole state at the critical obstacle strength.

\begin{acknowledgments}
This work was supported by the National Research Foundation of Korea (NRF-2018R1A2B3003373, NRF-2019M3E4A1080400) and the Institute for Basic Science in Korea (IBS-R009-D1).
\end{acknowledgments}

\appendix

\section{Vortex cluster shedding}\label{sec:Appendix1}

\begin{figure*}
    \centering
    \includegraphics[width=\linewidth]{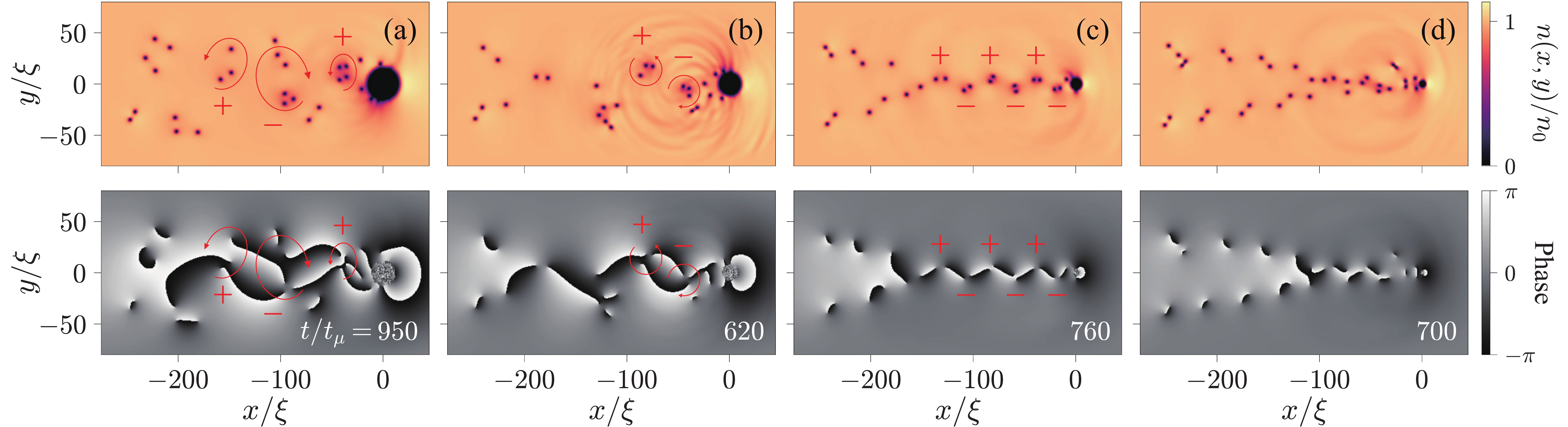}
    \caption{Vortex cluster shedding for strong impenetrable obstacles. The particle density (top) and phase (bottom) distributions of a BEC flowing past a strong impenetrable obstacle with $V_0/\mu=100$: (a) $\sigma/\xi=10$, (b) 7, (c) 3.5, and (d) 2. The obstacles were accelerated from $0.35c_s$ to $0.525c_s$ in (a), from $0.4c_s$ to $0.6c_s$ in (b,c), and from $0.45c_s$ to $0.575c_s$ in (d) for $1000t_\mu$. The circulation direction of each vortex cluster is indicated by the red arrow and sign.}
    \label{fig:FIGA1}
\end{figure*}

When an impenetrable obstacle moves at high velocity above $v_c$, it can generate vortex clusters consisting of many same-sign vortices~\cite{Reeves} due to its ability to pin multiple vortices. In Fig.~\ref{fig:FIGA1}, we present numerical results of the vortex cluster shedding for various obstacle sizes. In the numerical simulations, small Gaussian noises are added to the initial wave function of the BEC \cite{Von_Karman}, which break the left-right symmetry of the system with respect to the obstacle moving direction and facilitate the alternate shedding of vortex clusters with different signs of net circulation via hydrodynamic instability. The obstacle is accelerated up to above $0.5 c_s$. The size of the vortex clusters decreases with decreasing obstacle size. For obstacles with further reduced radii of $\sigma/\xi_s<4$, von Kármán streets of same-signed vortex pairs are observed [Figs.~\hyperref[fig:FIGA1]{\ref{fig:FIGA1}(c)} and \hyperref[fig:FIGA1]{\ref{fig:FIGA1}(d)}]~\cite{Von_Karman}. In Fig. \hyperref[fig:FIGA1]{\ref{fig:FIGA1}(d)} for our smallest obstacle,
we observe that the distances between the vortex clusters are reduced and the Kármán street structure collapses due to the interactions between adjacent clusters.


\begin{references}
\bibitem{footnote1}
For real-time evolution, we use a spatial domain of $(L_x, L_y) = (400, 200)\xi$ with $(N_x,N_y)=(600,300)$ equally distributed grids to simulate the periodic vortex shedding. The time interval is set by $\Delta t/t_{\mu}=0.04$.
\bibitem{footnote2}
In the velocity field calculation, if two adjacent coordinates have a phase difference above 85\% of $2\pi$, we subtract $2\pi$ from the original phase difference so to realize the inversion of the flow direction.
\end{references}
\end{document}